\documentclass[conference]{IEEEtran}
\IEEEoverridecommandlockouts

\def\BibTeX{{\rm B\kern-.05em{\sc i\kern-.025em b}\kern-.08em
    T\kern-.1667em\lower.7ex\hbox{E}\kern-.125emX}}
\usepackage[utf8]{inputenc}
\usepackage[english]{babel}
\usepackage{cite}
\usepackage{amsmath,amssymb,amsfonts}
\usepackage{algorithmic}
\usepackage{graphicx}
\usepackage{textcomp}
\usepackage{xcolor}
\usepackage[a4paper, total={184mm,239mm}]{geometry}
\usepackage[per-mode=symbol,per-symbol =/]{siunitx}
\usepackage[acronym]{glossaries}
\usepackage[capitalise]{cleveref}
\usepackage{float}
\usepackage{multirow}
\usepackage{array}
\usepackage{setspace}
\usepackage{enumitem}
\usepackage{url}
\usepackage{hhline}
\usepackage[font=footnotesize,labelfont=bf]{caption}
\ifCLASSOPTIONcompsoc
\usepackage[caption=false, font=normalsize, labelfont=sf, textfont=sf]{subfig}
\else
\usepackage[caption=false, font=footnotesize]{subfig}
\fi

\usepackage{amssymb}
\usepackage{pifont}
\DeclareSIUnit\GE{GE}
\DeclareSIUnit\kGE{\kilo\GE}
\DeclareSIUnit\MGE{\mega\GE}

\begin{document}

\title{Towards a RISC-V Open Platform for Next-generation Automotive ECUs\\
\thanks{This project has received funding from the European Union’s Horizon 2020 research and innovation programme under grant agreement No 871669.}
}

\author{
    \IEEEauthorblockN{%
    Luca Cuomo\textsuperscript{\textdagger}, %
    Claudio Scordino\textsuperscript{\textdagger}, %
    Alessandro Ottaviano\textsuperscript{\textasteriskcentered}, %
    Nils Wistoff\textsuperscript{\textasteriskcentered}, %
    Robert Balas\textsuperscript{\textasteriskcentered}, \\%
    Luca Benini\textsuperscript{\textasteriskcentered}, %
    Errico Guidieri\textsuperscript{\textdagger}, %
    Ida Maria Savino\textsuperscript{\textdagger} %
    }
    \IEEEauthorblockA{
        \textasteriskcentered~\textit{Integrated Systems Laboratory, ETH Zurich}, Switzerland \\
        \textdagger~\textit{Huawei Research Center}, Pisa, Italy \\
        \{l.cuomo,c.scordino,e.guidieri,i.savino\}@huawei.com \\
        \{aottaviano,nwistoff,balasr,lbenini\}@ethz.ch
    }
}

\maketitle

\begin{abstract}
The complexity of automotive systems is increasing quickly due
to the integration of novel functionalities such as assisted or autonomous driving.
However, increasing complexity poses considerable challenges to the automotive supply chain since the continuous addition of new hardware and network cabling is not considered tenable.
The availability of modern heterogeneous multi-processor chips represents a unique opportunity to reduce vehicle costs by integrating multiple functionalities into fewer Electronic Control Units (ECUs). In addition, the recent improvements in open-hardware technology allow to 
further reduce costs by avoiding lock-in solutions.

This paper presents a mixed-criticality multi-OS architecture for 
automotive ECUs based on open hardware and open-source technologies.
Safety-critical functionalities are executed by an AUTOSAR OS running on a RISC-V processor, while the Linux OS executes more advanced functionalities on a multi-core ARM CPU.
Besides presenting the implemented stack and the communication infrastructure, this paper provides a quantitative gap analysis between an HW/SW optimized version of the RISC-V processor and a COTS Arm Cortex-R in terms of real-time features, confirming that RISC-V is a valuable candidate for running AUTOSAR Classic stacks of next-generation automotive MCUs.
\end{abstract}


\section{Introduction}\label{sec:introduction}
For decades, automotive has been a very conservative industry, with software
functionalities operated by simple Electronic Control Units (ECUs) communicating
through domain-specific networks (e.g. CAN, LIN, FlexRay). 
However, the recent increase in the complexity of automotive systems due to the integration of novel functionalities, such as assisted or autonomous driving, poses considerable challenges to this industry.  
Modern luxury cars already contain more than 100 different ECUs~\cite{strandberg2018},
and the addition of new hardware has become an untenable task due to the amount of cabling inside the vehicle~\cite{mckinsey} and increasing space, weight, power and cost (SWaP-C).
Thus, nowadays, a significant opportunity for this industry is represented by the availability of asymmetric multi-processor (AMP) chips, which integrate (i) high-performance, multi-core application-class CPUs running a general-purpose OS (GPOS) such as Linux, (ii) slow real-time microcontrollers (MCUs) running a real-time operating system (RTOS),  and (iii) domain-specific accelerators.
The different processing units, in fact, could be used to integrate and consolidate multiple functionalities (even with different non-functional requirements) on the same ECU, reducing the amount of hardware and cabling inside the vehicle. 

In parallel to this trend, another interesting opportunity for the automotive industry is represented by the open hardware initiatives,  which aim at designing open instruction-set architectures (ISAs) to avoid vendor lock-in solutions and thus further reduce the recurrent costs faced by the OEMs.
In particular, the RISC-V ISA~\cite{riscv} is getting momentum across various industry domains as the future \textit{lingua franca} for computing and is widely considered a promising technology with significant potential also for the transportation domain~\cite{riscv-transportation}.

While previous works combining multi-OS and open hardware architectures for the automotive and space domains have been proposed, they often rely on symmetric/heterogeneous multi-processor (SMP and HMP) chips demanding hypervisor support for multi-OS execution, closed-source hardware architectures, and bespoke software libraries for intra-OS communications (Sec. \ref{sec:related-work}).  
This paper proposes a hardware and software stack for the automotive domain that leverages both AMP and RISC-V-based hardware towards the design of an open platform for automotive that relies on typical middleware employed in automotive.
In particular, the paper provides the following contributions:

\begin{itemize}
    \item We conceptualize a heterogeneous mixed-criticality system (MCS) with multi-OS architecture where a Linux-capable commercial multi-core system is paired with an open-source RISC-V MCU~\footnote{\url{https://github.com/pulp-platform/cheshire}} designed around CVA6~\cite{ariane} that runs an RTOS tailored for automotive, ERIKA Enterprise~\cite{erika}.
    \item We demonstrate the MCS system on a heterogeneous FPGA, namely the Xilinx Zynq Ultrascale+, which combines a \textit{hard macro} implementing an ARM-based multi-core system with programmable hardware implementing the RISC-V real-time MCU (Sec. \ref{sec:architecture}).
    To the best of the authors' knowledge, this is the first work that attempts to adopt a multi-OS open-source stack for automotive based on open hardware.
    \item We conduct a quantitative gap analysis of the CVA6 \mbox{RISC-V} MCU against an Arm-based real-time MCU (Cortex-R series) available on the heterogeneous FPGA in terms of interrupt response time, showing a significant performance gap of the RISC-V interrupt support as intended in the Privileged specifications~\cite{RISCV_II}.  
    \item We extend the real-time capabilities of the RISC-V CVA6 MCU by coupling the core with a RISC-V fast interrupt controller (CLIC~\cite{clic}), which allows achieving competitive real-time performance against the Arm competitor, paving the road for further development of the RISC-V ISA in the transportation domain (Sec. \ref{sec:experiments}).
\end{itemize}



\section{Related work}~\label{sec:related-work}

\paragraph{Automotive operating systems}
In the '90s some German and French companies joined their efforts to create the OSEK/VDX consortium~\cite{osek}, aiming at creating an open standard for the operating system and the communication stack of automotive embedded systems. Some parts of these specifications were then standardized in ISO~17356~\cite{iso17356}. Some open-source implementations have been 
proposed during the years, with the most notable projects being Trampoline~\cite{trampoline}
and ERIKA Enterprise~\cite{erika}.

The AUTomotive Open System ARchitecture (AUTOSAR) consortium, started in 2004, has coordinated and driven a standardization effort in the last two decades to handle the growing complexity of the software inside vehicles.
The specification (namely, AUTOSAR Classic~\cite{autosar-cp}) extended the original OSEK/VDX standard to design the stack for simple automotive ECUs executing tiny real-time operating  systems (RTOSs) and communicating through domain-specific networks (e.g. CAN, LIN, FlexRay).
The advent of modern functionalities, like assisted or autonomous driving, has
then forced the consortium to release in 2017 an additional specification (namely, AUTOSAR Adaptive~\cite{autosar-ap}) for a more dynamic platform based on the POSIX API~\cite{posix}
and also capable of High-Performance Computing (HPC). The consortium has also provided an
exemplary implementation of part of the specification running on the Linux OS.

The idea of using a Multi-OS architecture for automotive is well understood in the literature. 
Burgio et al.~\cite{burgio2017} proposed a Multi-OS architecture developed in the context of the HERCULES European project.
Despite employing the same RTOS presented in this paper (i.e., ERIKA Enterprise),
they relied on a closed-source Heterogeneous Multi-Processor (HMP) architecture based on the Arm \textit{big.Little} concept, and run multiple operating systems on top of an open-source hypervisor.
Moreover, the communication between the different operating systems was achieved through ad-hoc
libraries rather than using standard middleware employed in automotive.

\begin{figure}
\centering
\includegraphics[width=0.9\linewidth]{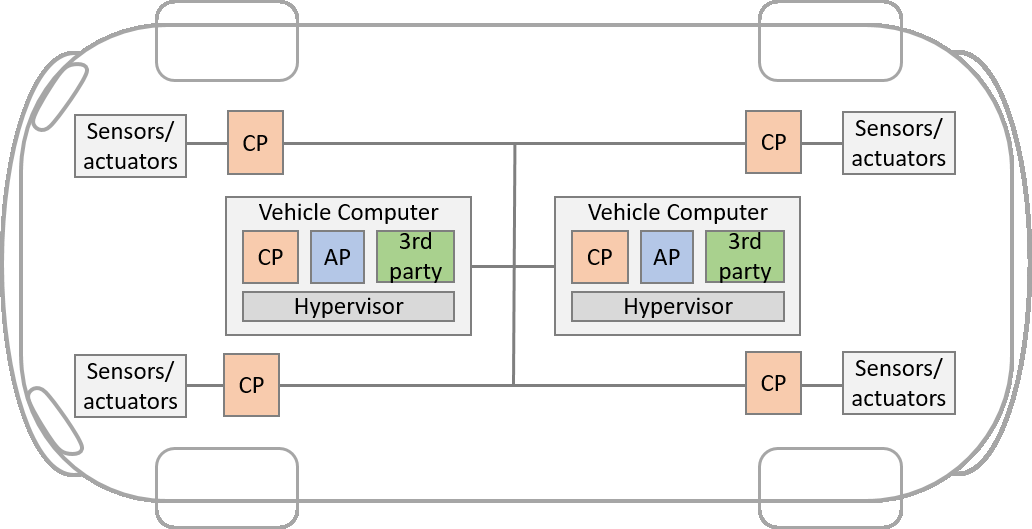}
\caption{Example of zonal architecture.}
\label{fig:zone-architecture}
\end{figure}

On the industrial side, silicon vendors have started designing AMP system-on-chips (SoC) comprising  
a high-end multi-core processor (possibly in a \textit{big.Little} configuration) tasked to run the GPOS, and a slower microcontroller tasked to run a safety-critical RTOS. 
An example is the i.MX8 chip by NXP~\cite{nxp_imx8}, which includes Arm A52 and A72 cores along with Arm M4 cores. The intra-OS communication is delegated to bidirectional connection-less Remote Processor Messaging (RPM) interfaces.
More recently, Arm has proposed the first high-performance 64-bit real-time processor of the R-series, Cortex-R82. Despite the announced use case as a storage controller for the IoT domain, such a processor is a candidate for future dual GPOS/RTOS execution on a common hardware platform.
In fact, automotive OEMs are already transitioning from a domain architecture to a zonal architecture~\cite{alparslan2021next} similar to the one shown in Figure~\ref{fig:zone-architecture}, where few Vehicle Computers run a multi-domain stack that includes both AUTOSAR Classic Platform (CP) and AUTOSAR Adaptive Platform (AP), along with 3rd party software (e.g., ROS2, plain Linux, other operating systems, etc.).

Albeit several architectural similarities between the available multi-OS platforms, this work distinguishes itself on multiple angles: (i) it relies on a fully open-source real-time MCU developed within the ever-growing RISC-V ecosystem to handle safety-critical tasks, (ii) the automotive software stack running on RISC-V is based on an open-source RTOS~\cite{erika}, and (iii) it enables RISC-V as the leading actor of the transition towards zonal architectures, that is currently not taking into account open hardware processors.

The following paragraph analyzes the state-of-the-art in bringing RISC-V architectures into the automotive domain in the last few years.

\paragraph{Automotive RISC-V architectures}
RISC-V technology has received much attention during the last years after 
some vendors released high-end multi-core CPUs at 1.5~GHz  
capable of running Linux.
Some recent work has also been devoted to increasing the safety levels of RISC-V architectures.
De-RISC~\cite{derisc, derisc2} is an H2020 project aiming at designing a RISC-V processor and 
software stack for safety-critical systems. However, the project is mainly focused 
on the space industry.

Abella et al.~\cite{abella2021security} identified some issues of the RISC-V 
ecosystem related to security and reliability and provided four contributions
to implement lock-step and system-level testing. Their work is very relevant 
for the automotive domain and could be implemented on top of our proposed
architecture.

Pietzsch~\cite{pietzsch2021risc} presented EMSA5-FS, a 32-bit, single-issue, in-order, 5-stage RISC-V processor specifically designed for functional safety.
Cosimi et al.~\cite{cosimi2022} proposed to mix core independent peripherals, including a Performance  Monitoring Unit (PMU), an Error Management Unit, and an Execution Tracing Unit, to increase  the safety integrity level of an application running on a RISC-V platform
up to the highest automotive level (i.e., ASIL-D).
Their implementation has been done on the same evaluation board used in our experiments 
(i.e. Xilinx ZCU102~\cite{zcu102}) and can therefore be fully integrated with 
the architecture presented in this paper.

Quite recently, Gruin et al.~\cite{Gruin2021} presented MINOTAuR, a 
timing predictable open source RISC-V core, based on the same
hardware architecture used in this work.
The experimental results have shown an overhead of 10\% compared to
the unmodified core, obtained through partial speculative execution.
Although their work does not explicitly address automotive,
predictability is a non-functional requirement needed at every time- and safety-critical
domain (including automotive).

Very recently, SiFive and Renesas have announced a long-term
collaboration to design and produce RISC-V processors for the automotive
domain~\cite{sifive-auto}.
These ISO26262-qualified processors will all have the same ISA to increase code
portability.

This work relies on the open-source 64-bit core CVA6~\cite{ariane}. It extends its real-time capabilities to serve as a time- and safety-critical RISC-V system in a multi-OS platform, closing the gap with existing embedded COTS solutions (Sec. \ref{sec:experiments}).

\paragraph{Automotive communication protocols}
To ensure a reasonable and manageable complexity through composability,
the automotive industry is replacing the original
signal-oriented communication with modern service-oriented
architectures (SoA).
Using this paradigm, the various software components are decoupled from each other 
and communicate by requesting and providing "services".
Each component can be designed in isolation, and the system is assembled 
by composing and integrating the various functionalities.

Proposed initially by BMW, Scalable service-Oriented
MiddlewarE over IP (SOME/IP)~\cite{some-ip} is a SoA protocol specifically
designed for Ethernet-based communications in automotive.
This standard specifies the serialization mechanism,
the service discovery and the integration with the AUTOSAR
stack. 

More recently, Data Distribution Service (DDS)~\cite{dds-specs}
started attracting a growing interest from the automotive industry~\cite{rti-dds}.
Originally proposed in 2001, DDS became an Object Management Group (OMG)
standard in 2004, with several open-source implementations available
nowadays. The DDS specifications~\cite{dds-specs} describe a \emph{Data-Centric Publish-Subscribe}
model for distributed application communication. This model builds on
the concept of a ``global data space'' contributed by publishers and
accessed by subscribers: each time a publisher posts new data into this
global data space, the DDS middleware propagates the information to all
interested subscribers. 
The data-centric communication allows the decoupling of publishers from
subscribers, thus building a very scalable and flexible architecture.
The underlying \emph{data model} specifies the set of data items,
identified by ``topics``. 

Nowadays, DDS is natively supported by most frameworks used in automotive
--- namely, AUTOSAR Classic~\cite{Scordino2022}, AUTOSAR Adaptive and ROS~\cite{ros}.
Note that, according to some recent investigations~\cite{pohnl2022middleware}, 
the ROS framework is already  being used by about 80\% of the automotive OEMs 
and Tier-1s developing autonomous vehicles.

\section{System architecture}~\label{sec:architecture}

\begin{figure}[tb]
	\centering
	\subfloat[System architecture overview. The figure highlights the multi-core system of application-class cores tasked to run the GPOS (left-hand side) and the single-core RISC-V MCU system tasked to run the RTOS (right-hand side). Inter-OS communication is based on DDS at the software level and relies on AXI4 as the hardware transport protocol.]{%
       \includegraphics[width=0.8\columnwidth]{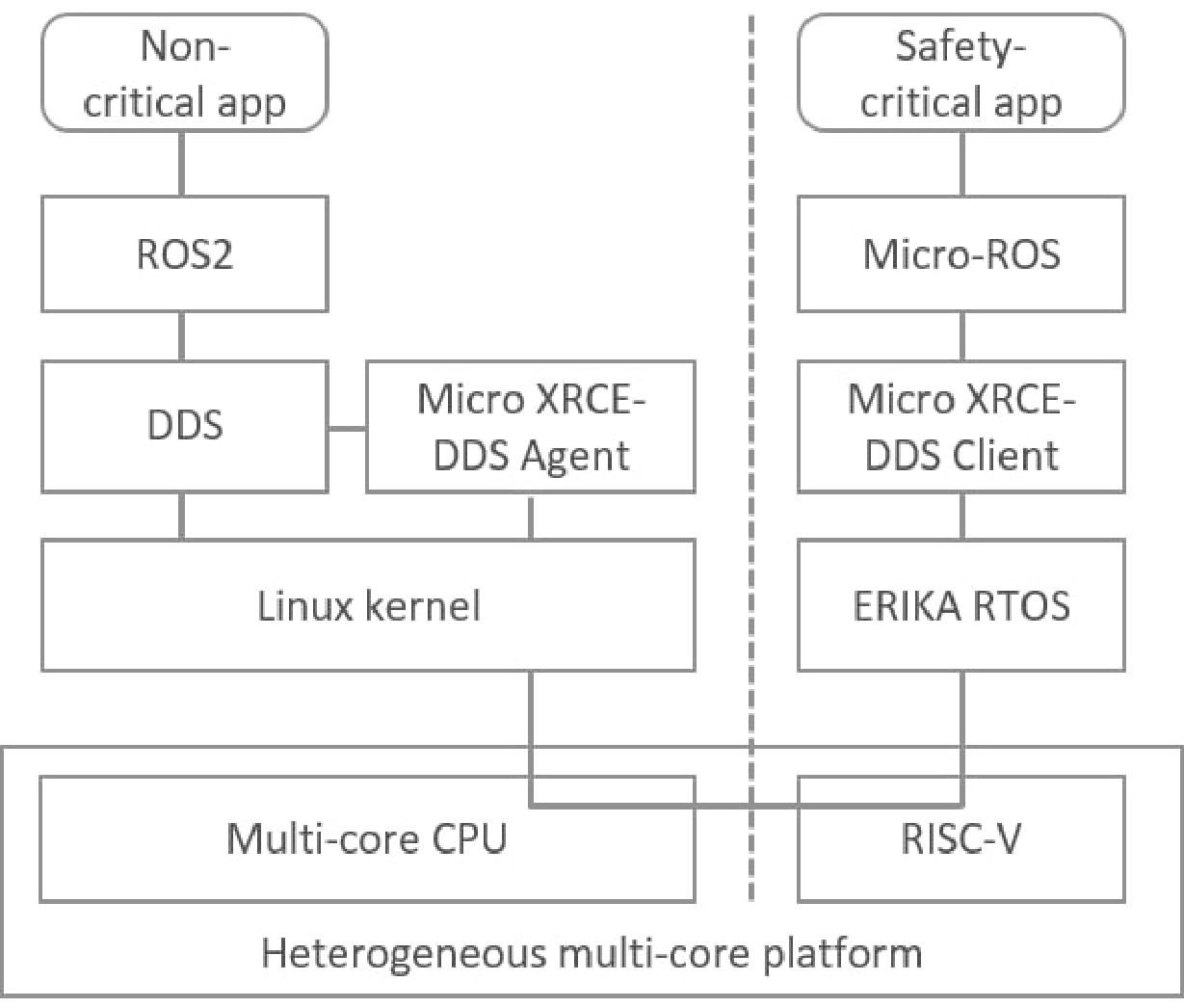}
       \label{fig:architecture}
       }
       
    \subfloat[CVA6 RISC-V MCU architecture overview (simplified). The figure highlights the main building blocks of the system and its AXI4-based interface towards the multi-core domain.]{\includegraphics[width=0.8\columnwidth]{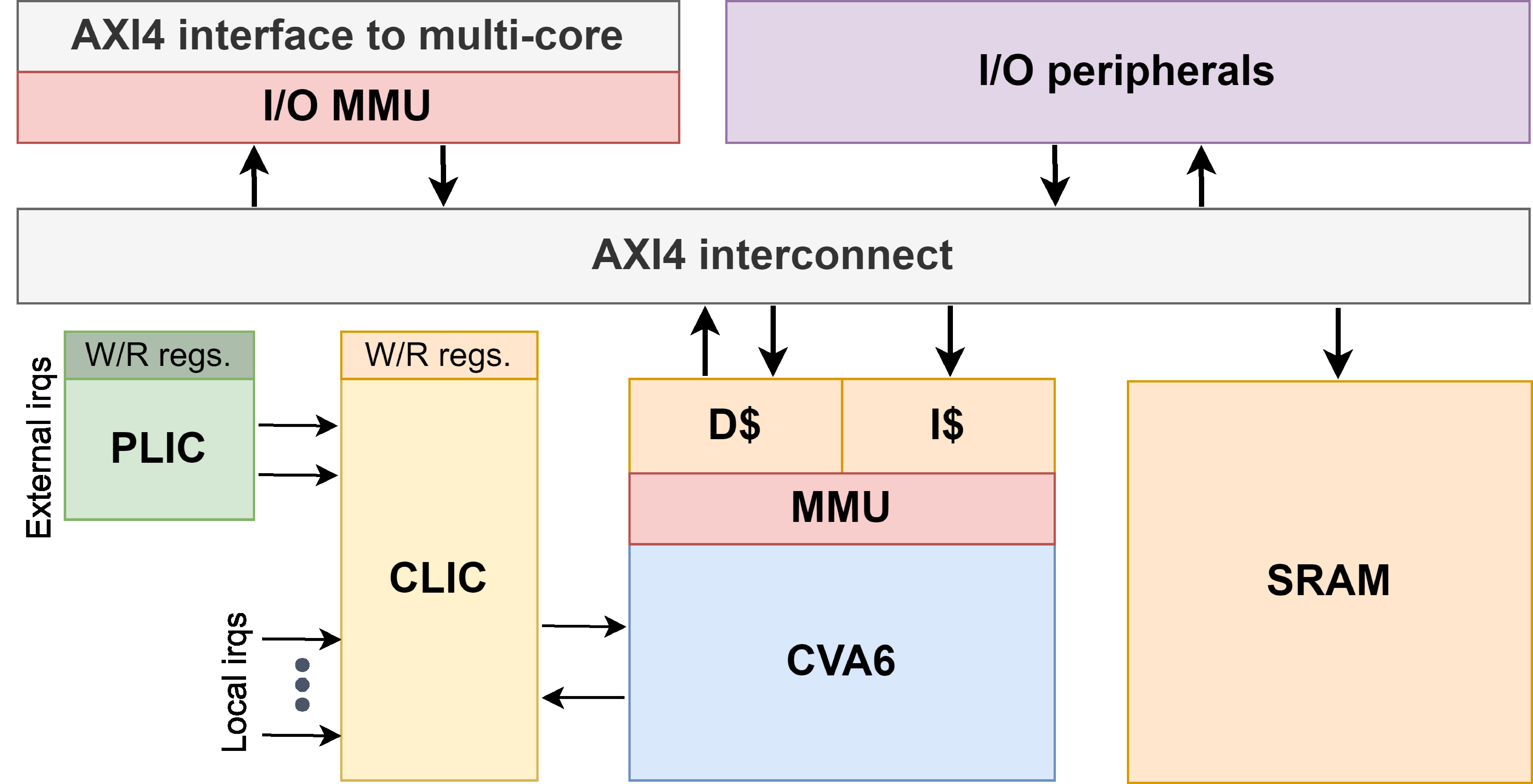}
        \label{fig:cva6_mcu}
        }
    \caption{Overview of the heterogeneous automotive architecture and highlight on the CVA6 RISC-V MCU.}
\end{figure}

As shown in Figure~\ref{fig:architecture}, the proposed mixed-criticality architecture 
consists of an AMP system-on-chip (SoC) comprising  
a high-end multi-core processor tasked to run the GPOS, and a slower microcontroller tasked to run a safety-critical RTOS. 

We design the RISC-V MCU around the 64-bit CVA6 \mbox{RISC-V} core. 
CVA6 is a 6-stage, single-issue, in-order core implementing the G and C extensions of the 64-bit RISC-V instruction set (RV64GC). 
The core implements a Translation Lookaside Buffer (TLB) to accelerate address translations from the virtual to the physical domain and a classic branch predictor consisting of a branch target buffer (BTB), a branch history table (BHT), and a return address stack (RAS). The core employed in this work is configured with a 32-kiB write-through L1 data cache and a  16-kiB instruction cache.
Besides the core, the MCU hosts 128-kiB scratchpad memory (SPM), a direct memory access (DMA) engine, and low-latency peripherals (SPI, I2C, UART) for off-chip communication.
The MCU relies on an AXI4-compliant, on-chip, non-coherent interconnect system. AXI4 interfaces are exposed to the multi-core domain through a software-managed IOMMU (such as in~\cite{akurth_herov2}) consisting of an IO translation lookaside buffer (IOTLB) to efficiently translate virtual user-space application addresses from the multi-core domain to physical memory.
Fig. \ref{fig:cva6_mcu} depicts the RISC-V MCU and its hardware interface towards the application-class host.

In the embedded domain, a general-purpose core's real-time capabilities strongly depend on its interrupt controller's design. This is a crucial and functional requirement in safety- and time-critical systems such as those operating in the automotive domain, aiming at minimizing interrupt latency and context switch time.
First, CVA6 lacks support for \textit{vectored interrupts}, which store the interrupt service routine of each interrupt at a separate address. 
Albeit increasing the code size as the vector table's size grows, this mechanism helps reduce the overall interrupt response time.
Furthermore, CVA6's native interrupt architecture consists of classic RISC-V PLIC and CLINT controllers from the RISC-V Privileged Specifications~\cite{RISCV_II}. The core hosts three level sensitive interrupt signals: machine-mode timer interrupt, machine-mode software interrupt (inter-processor interrupt), and machine-\slash supervisor-mode external interrupts, respectively.
The machine timer and machine software interrupt pending registers --- \texttt{mtip} and \texttt{msip} respectively --- are provided by a Core Local Interruptor (CLINT) hardware Intellectual Property (IP), which generates one interrupt for each hardware thread (\textit{hart}, a RISC-V execution context). While \texttt{mtip} generates timer interrupts with a specific frequency, \texttt{msip} handles communications among processors by interrupting harts on writes/reads of dedicated memory-mapped registers.
The first 12 interrupts' identifiers are reserved for timer, software, and external interrupts in the machine (M), supervisor (S), and user (U) privilege modes. Other interrupt entries up to XLEN (for an RV64 processor such as CVA6, XLEN=64) are platform specific and referred to as \textit{local} interrupts~\cite{RISCV_II}.
Finally, the machine external and supervisor external interrupt pending registers \texttt{meip/seip} bring the information from \textit{external} devices to the \textit{hart}.
The Platform Local Interrupt Controller (PLIC)~\cite{plic} provides centralized interrupt prioritization and routes shared platform-level interrupts among multiple harts via the \texttt{meip\slash seip} interrupt signals. The PLIC does not support interrupt preemption (nesting), nor runtime-configurable interrupt priorities and interrupt threshold control, which must be simulated in software.
As highlighted in Sec. \ref{sec:introduction} and further detailed in Sec. \ref{sec:experiments}, such native features are insufficient to fulfill functional real-time requirements.
An essential contribution of this work is the enhancement of CVA6's real-time capabilities in terms of interrupt response to achieve a competitive advantage against existing COTS real-time MCUs.

To ease the design and development of the AMP system, the RISC-V MCU hosting the RTOS has been implemented on a heterogeneous Xilinx Zynq Ultrascale+ FPGA~\cite{zcu102} as part of the Programmable Logic (PL), thus taking advantage of the existing multi-core COTS SoC (Processing System, PS) to host the GPOS.
The PS consists of an industry-standard, quad-core, 64-bit Armv8 Cortex-A53 application-class core featuring 32 KiB L1 instruction and data cache per core and a 1 MiB L2 cache shared by all four cores and clocked at 1.2 GHz and a dual-core Cortex-R5F \textit{real time unit}.
The Arm Cortex-R is employed to conduct the performance gap analysis with real-time enhanced CVA6 despite the difference in XLEN (32-bit and 64-bit, respectively). 
The CVA6 MCU has been synthesized on the PL targeting 50~MHz frequency. The following describes the software stacks running on the various processors.


\subsection{Real-time OS}

The OSEK/VDX and AUTOSAR Classic standards specify the design of a tiny RTOS for automotive.
The programming paradigm is ``run-to-completion,'' and the configuration (e.g. number of tasks) is statically defined at compile time.
In this type of operating system, the Interrupt Service Routines (ISR) 
are divided into two categories:
\begin{itemize}
    \item ISR1: High-priority low-overhead routines that cannot call syscalls;
    \item ISR2: Priority-based routines, which could imply a rescheduling once finished.
\end{itemize}
In the proposed architecture, we have used the ERIKA Enterprise RTOS~\cite{erika}.
This open-source RTOS supports various microcontroller architectures and
is used in several European research projects and industrial automotive products. 
A fork of ERIKA Enterprise recently received ISO26262 ASIL-D qualification
(the highest safety level for automotive). 
Moreover, there is an ongoing discussion with the AUTOSAR consortium to release this RTOS 
within the Classic demonstrator under the name of ``Open-ERIKA''~\cite{gai2022}.
In the context of the AMPERE H2020 project, the ERIKA RTOS has been ported
and executed on the RISC-V architecture.

\subsection{General-purpose OS}
Linux is a well-known operating system implementing the POSIX API.
Its performance, open-source license, and portability made it a perfect candidate for the general-purpose operating system running on the multi-core ARM PS.

During the last decades, several attempts have been made to improve the real-time performance
of Linux systems~\cite{scordino2006linux}. From time to time, some support (e.g. preemptible kernel, priority inheritance protocol, high-resolution timers, SCHED\_DEADLINE real-time scheduler~\cite{lelli2016}) have been merged in the official kernel.
PREEMPT\_RT~\cite{preempt-rt} is a long-term project sponsored by the Linux Foundation to improve the real-time capabilities of the operating system. The primary outcome of this project is a kernel patch that reduces the maximum latency experienced by applications and is expected to be merged in the mainline ``Vanilla'' codebase.
To improve the overall responsiveness of the proposed platform, we have therefore
re-compiled the Linux kernel applying the PREEMPT\_RT patch and enabling the maximum preemption level.

It is worth mentioning the existence of a joint initiative, ELISA~\cite{ELISA}, aiming
at easing the certification of this operating system in safety-critical environments. In the automotive scenario, the ELISA project aims to reach the ASIL-B certification of the OS. However, the project has not yet provided a process to obtain such a qualification.

\subsection{Intra-OS communication}

According to the latest trends in automotive~\cite{Scordino2022}, the inter- and intra-OS communications have been entirely based on the DDS standard.
The intra-OS communication between processes running on the Linux OS has been implemented through an open-source DDS middleware (i.e., Fast-DDS, formerly known as Fast-RTPS).
The inter-OS communication between Linux and ERIKA, instead, has been based on DDS-XRCE~\cite{dds-xrce}, a DDS protocol specifically designed by OMG 
for resource-constrained systems. 
As shown in Figure~\ref{fig:architecture}, in this client-server protocol, the devices (clients) communicate with an XRCE Agent (server), which provides the intermediate bridging service towards the DDS Data Global Space. 
In particular, we have integrated eProsima's Micro XRCE-DDS stack~\cite{micro-XRCE-DDS} 
(part of the Micro-ROS project~\cite{micro-ROS}) on the ERIKA RTOS.

\section{Evaluation}~\label{sec:experiments}

In this section, we analyze and characterize the proposed automotive platform in terms of real-time capabilities, focusing on interrupt handling latency on both the multi-core system running the GPOS and the MCU running the RTOS, as well as the inter-domain communication time overhead:

\begin{itemize}
    \item We analyze real-time extensions of the Linux kernel to suit the automotive domain better.
    \item We characterize the middleware layer for intra-OS communication.
    \item We optimize the real-time CVA6 MCU in hardware to boost its interrupt response capabilities with the integration of the RISC-V CLIC as the central interrupt controller for CVA6 and conduct a performance gap analysis with the COTS Arm Cortex-R5 already available in the PS of the FPGA.
\end{itemize}

\subsection{Non-critical multi-core domain: Linux GPOS}~\label{sec:experiments-linux}
For Linux, we have used an Ubuntu filesystem and the Foxy version of ROS2 on top of Fast-DDS.  
The Linux kernel was version 4.19, patched with the PREEMPT\_RT patch. 
We have then run a set of tests 
to measure the latency introduced by the OS. 
The system has been stressed by creating interference through both the {\tt find} command
(generating I/O traffic by scanning the filesystem on the SD memory and printing on the console) and
through the {\tt stress} program generating CPU, memory and I/O interference: ./rt-test/stress -c 8 -i 8 -m 8 --vm-bytes 
8000000.

The worst-case latency has been measured through the {\tt cyclictest} tool provided
by the Linux kernel community developing the PREEMPT\_RT patch. The tool has been run
with the following options: 
./cyclictest --mlockall --smp --priority=80
    --interval=200 --distance=0 --duration=5m.

The experimental results have shown a worst-case latency of 13.4~ms without PREEMPT\_RT
and~\textbf{159~$\mu$s} with PREEMPT\_RT. This means that the maximum latency experienced by user-level applications has been reduced of about~\textbf{99\%} by simply applying the patch and recompiling the Linux kernel.

\subsection{Inter-domain communication}~\label{sec:experiments-dds}


The communication latency has been evaluated through a ``ping-pong'' application
that measured the round-trip time from Linux to ERIKA and back to Linux.
The involved processes on Linux (i.e., DDS Agent and ROS2 application) have been
scheduled using a real-time priority (i.e., SCHED\_RR with priority 99). 
Data has been exchanged through a non-cached shared memory area. We selected UART as the interrupt source, the only visible from both operating systems. 
The experimental results showed a minimum, average and maximum communication time
of \textbf{2.0}, \textbf{2.2}, and \textbf{3.7 msec}, respectively.
It is essential to highlight that the Micro-ROS framework has a periodic engine
which added some delay to the communication.
In particular, the {\tt clc\_executor\_spin\_some} function had a period of 1 msec,
while all the other interactions were event-driven.

\subsection{Safety critical RISC-V MCU domain: ERIKA RTOS}~\label{sec:experiments-rtos}

When porting the ERIKA Enterprise RTOS on RISC-V, we have taken 
inspiration from the previous FreeRTOS optimization~\cite{balas2021risc}.
We have initially optimized interrupt handling by emulating the local 
\textit{interrupt levels} through an array statically generated by the OS tools since the core does not natively support them. 

The performance of the ERIKA RTOS when running on
RISC-V has been measured through an existing benchmark~\cite{scordino2020real} 
that measures the time
needed by the RTOS for performing a set of critical scheduling
activities (e.g., task activation time, task exit time, ISR call time,
etc.).
The test suite also allows benchmarking the latency of the two types of interrupt
service routines available in AUTOSAR Classic kernels (i.e. ISR1 and ISR2)
that have been previously illustrated.
The tested functions are namely:
\begin{itemize}
\item {\tt act}: activates a higher priority task and measures how long
it takes to start its execution.
\item {\tt actl}: activates a low-priority task and measures how long it
takes to return to the caller.
\item {\tt intdisable}: measures the time needed for disabling all interrupts.
\item {\tt intenable}: measures the time needed for enabling all interrupts.
\item {\tt isrentry}: measures the time elapsed between the occurrence
of an interrupt and the execution of the related ISR1 handler.
\item {\tt isr2entry}: measures the time elapsed between the occurrence
of an interrupt and the execution of the related ISR2 handler.
\item {\tt isrexit} : measures the time elapsed between the end of an
interrupt handler and when the task previously running resumes execution.
\item {\tt istentry}: measures the time elapsed between the end of an
interrupt handler and the execution of the task activated by such
interrupt handler.
\item {\tt istexit}: measures the time elapsed between the end of a task
handling an interrupt and when the task previously running resumes
execution.
\item {\tt terml}: measures the time needed for terminating a task and
switching to a lower priority one.
\end{itemize}

Execution times have been measured in processor clock cycles through the \texttt{mcycle} {\tt CSR} register.
The same benchmark has been executed to evaluate the performance of the Cortex-R5 and the Cortex-A53 cores available on the ZCU102 board. For the Cortex-R5, the number of cycles has been measured through the {\tt PMCCNTR} register. 
For the Cortex-A53, instead, cycles have been measured through the cycle counter register {\tt PMCCNTR-EL0}: \_\_asm\_\_ \_\_volatile\_\_
    ("MRC p15, 0, \%0, c9, c13, 0" : "=r" (cycles)); 

It is essential to point out that, in the case of Cortex-A, the RTOS has been run on top of a hypervisor (namely, Jailhouse~\cite{jailhouse}) according to the typical configuration used when running RTOSs on Cortex-A processors. 
The presence of the underlying hypervisor, however, implied some 
non-negligible latency to trap and re-inject interrupts to the guest RTOS.
The possible interference from Linux on shared hardware resources has been removed by inducing the Linux kernel in panic mode through the following command: echo c > /proc/sysrq-trigger.

\begin{figure}[thb]
\includegraphics[width=\linewidth]{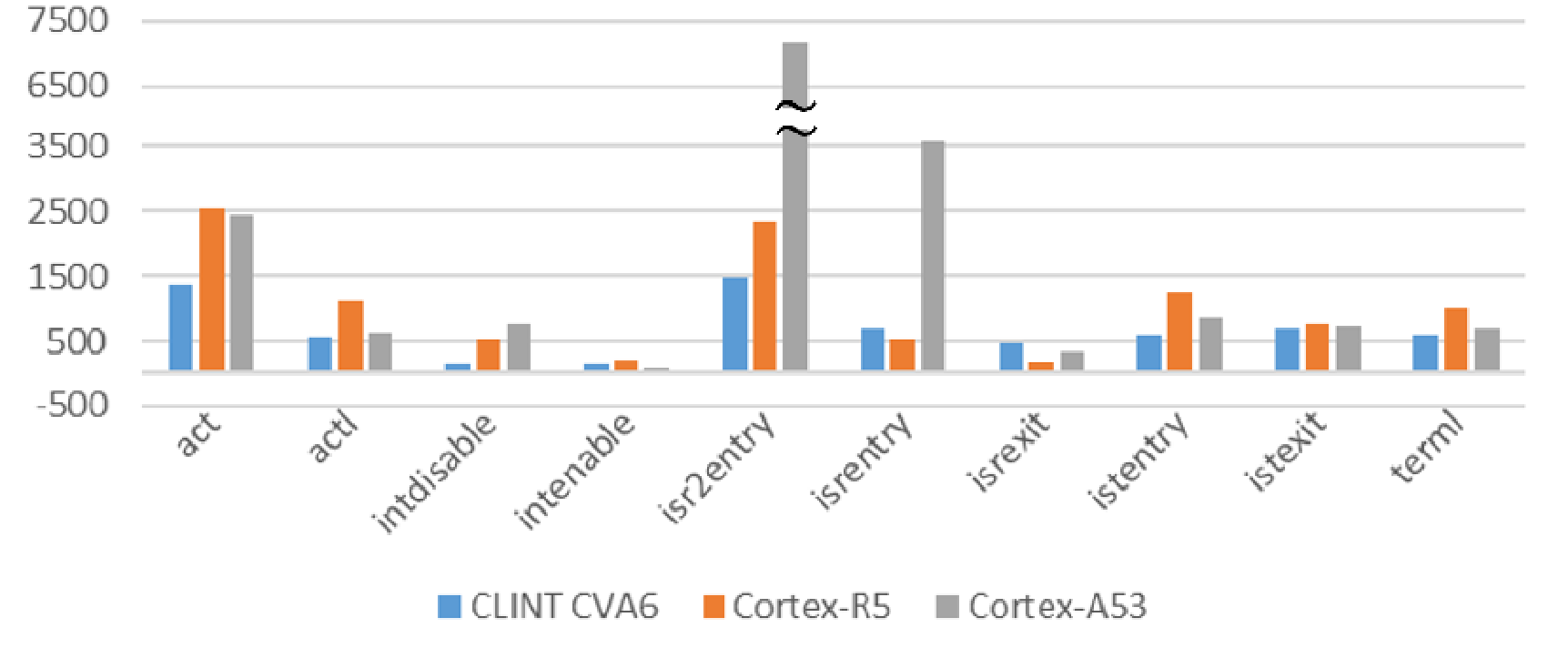}
\caption{RTOS performance (worst-case processor cycles).}
\label{fig:benchmarks}
\end{figure}

Since in safety-critical domains, such as automotive, we are most interested in bounding
the time needed for the various operations, we have restricted our analysis to the
worst-case number of cycles, measured over 100 consecutive runs.
The reported values in Fig.~\ref{fig:benchmarks} show that when CLINT/PLIC are being used, and no software optimization has been applied, the RISC-V soft-core can provide performance in the same order of magnitude of 
state of the art (i.e., Cortex-R5). In particular, the worst-case cycles are higher only for handling ISR1 interrupts.
The reported values also confirm the non-negligible latency introduced by the interrupt injection mechanism on the hypervisor on Cortex-A. This strengthens the benefits of designing a mixed-criticality architecture through an AMP SoC rather than a hypervisor-based approach on an SMP SoC.

\subsection{Software-driven RTOS optimization}~\label{sec:experiments-optimization}

The next step consisted in optimizing the code of the RTOS to obtain better performance in all the tested processors. The first optimization consisted in modifying the ISR2 handling by avoiding activating the ISR as a Task and directly calling the handler (i.e. not calling {\tt osEE\_activate\_isr2()}). Moreover, similarly to~\cite{balas2021risc}, we have used the {\tt -O3} optimization level of the GCC compiler. 

\begin{figure}[thb]
\includegraphics[width=\linewidth]{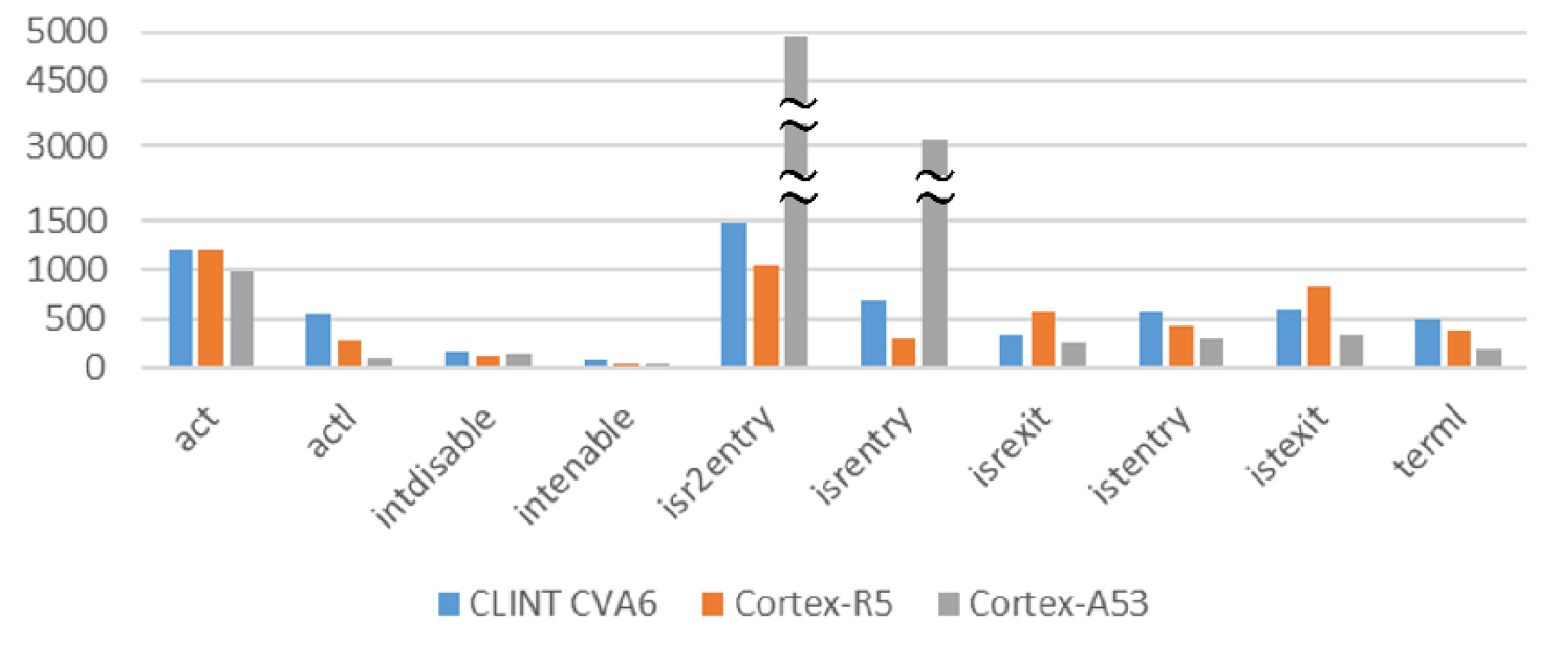}
\caption{RTOS optimized performance (worst-case processor cycles).}
\label{fig:optimization}
\end{figure}

Fig.~\ref{fig:optimization} reports the worst-case
number of cycles, still measured over 100 consecutive runs. 
The benefits of the optimizations can be appreciated across all the architectures. 
However, the proposed SW optimization of the RTOS on Cortex-R benefited more than the implementation on the RISC-V MCU, in some cases reducing the worst-case cycles to less than 25\% of the original value (e.g., in the {\tt actl}, {\tt intdisable} and {\tt intenable} tests). 

From the presented values, we can see that the selected RISC-V processor still shows lower performance in terms of interrupt latency than the competing ARM Cortex-R5 architecture.
As already discussed in Sec. \ref{sec:architecture}, we identify the bottleneck of the design in CVA6's interrupt handling support, which is not tuned for targeting fast-interrupt management and low interrupt latency, typically enabled through the following HW/SW mechanisms:

\begin{enumerate}
    \item Hardware support for fine-grained and configurable interrupt priorities
    \item Late-arriving interrupt behavior (preemption and nesting)~\cite{cortex-m}
    \item Context save/restore optimization with back-to-back interrupts (tail chaining)
    \item Banked stack pointer~\cite{cortex-m} (i.e. different stack pointers for different privilege levels)
    \item Hardware support for automatic saving of registers during the context switch.
\end{enumerate}

The following section addresses the first three of the above-mentioned design items. To this aim, we extend the current CLINT interrupt controller with a Core Local Interrupt Controller (CLIC)~\cite{clic} and evaluate the ERIKA RTOS's performance.
The remaining two design items involve implementing more advanced hardware features in the processor and will be investigated in future work.

\subsection{Hardware-driven real-time optimization: RISC-V CLIC fast interrupt controller}

\begin{figure*}[htb]
\includegraphics[width=\textwidth]{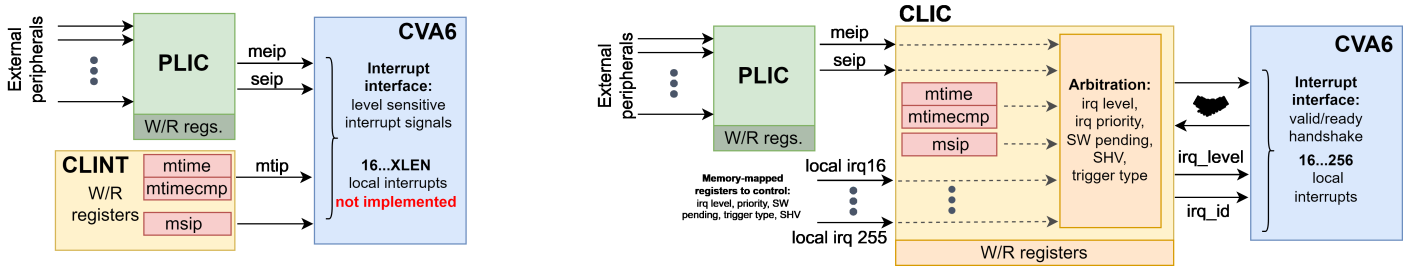}
\caption{\textbf{(a)} Original CLINT + PLIC interrupt interface.
    \textbf{(b)} Improved CLIC + PLIC interrupt interface.}
\label{fig:cva6_clic_plic}
\end{figure*}

As mentioned in Sec. \ref{sec:architecture}, both native CVA6 and its interrupt controller architecture need adaptations to fulfill real-time needs.
We first modify CVA6's interrupt interface by replacing level-sensitive interrupts with a handshake mechanism carrying the interrupt identifier and the request to the processor that acknowledges the handshake. We then add support for vectored interrupts by implementing an interrupt identifier decoding logic to compute the jumping address of the vector table.
In the second step, we extend the CLINT with the Core Local Interrupt Controller (CLIC).
We employ an open-source implementation of the CLIC\footnote{\url{https://github.com/pulp-platform/clic}} that reflects the latest status of the RISC-V CLIC draft specifications. The integration process includes the addition of specific CSRs registers in the processor's micro-architecture as from specifications~\cite{clic}.
The CLIC introduces several improvements to the standard CLINT to achieve faster interrupt handling. Among those are dedicated memory-mapped registers for software configurable interrupt priority and levels at the granularity of each interrupt line, runtime-configurable interrupt mode and trigger type, and support for interrupt preemption in the same privilege level.
Selective hardware vectoring enables the programmer to optimize each incoming interrupt for either faster response (vectored mode) or smaller code size (direct mode, when each interrupt traps to the same exception handler address).
Lastly, the CLIC introduces a novel CSR, namely \texttt{mnxti}~\cite{clic} to accelerate the handling of back-to-back interrupts, a phenomenon called \textit{tail-chaining}, which we have implemented in the CVA6 core.
CVA6 interrupt handling is modified as in Fig.~\ref{fig:cva6_clic_plic}.
In the improved design, the PLIC still arbitrates external system-level interrupts, and the legacy CLINT generates the timer interrupt. These interrupts are routed through the centralized CLIC interrupt source. Similarly, inter-processor interrupts are fired by writing to the corresponding CLIC memory-mapped registers.
Finally, local interrupts can be extended to 4096 lines instead of limited to the processor's \texttt{XLEN}. We implement 256 input interrupt lines arbitrated by the CLIC in this work.

\begin{figure}[htb]
\includegraphics[width=\linewidth]{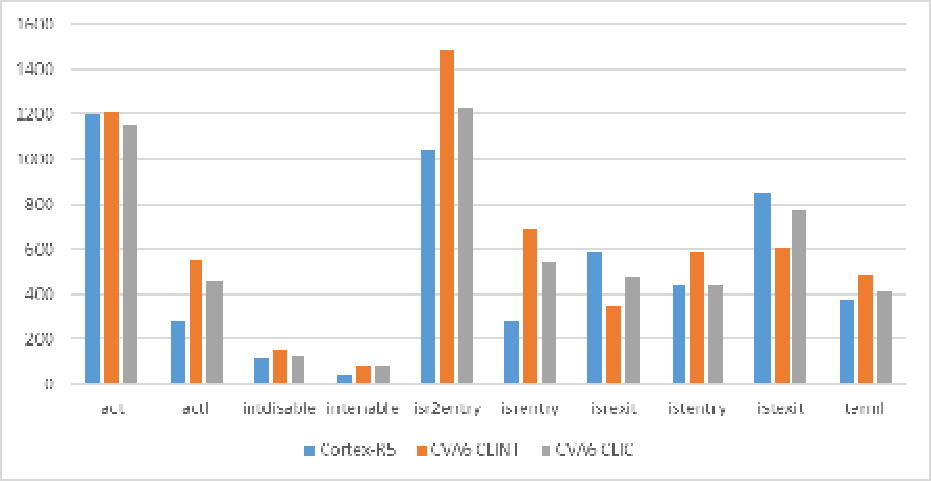}
\caption{Worst-case overhead of CVA6 with CLINT and CLIC interrupt controllers against the Cortex-R5.}
\label{fig:clint-clic-comparison}
\end{figure}

Fig.~\ref{fig:clint-clic-comparison}
shows the experimental results on the improved hardware architecture with {\tt GCC -O3} optimization level at compile time on the selected benchmark. We notice that the worst-case overhead on RISC-V has become closer to the one measured on
the competitor MCU (i.e. ARM Cortex-R5). 
The previous spikes (i.e. {\tt actl} and {\tt isrentry}) have been significantly reduced.
Moreover, for 4 metrics (i.e. {\tt act}, 
{\tt isrentry}, {\tt istentry} and {\tt istexit}) 
the number of cycles needed by the RTOS is equal to or even lower than on the Cortex-R5. 
These experimental results confirm that RISC-V is
a valuable technology for running AUTOSAR Classic stacks of next-generation automotive MCUs, and can be further improved to surpass closed-source commercial solutions.


\section{Conclusions}~\label{sec:conclusions}

In this paper, we have illustrated some trends and challenges occurring in the automotive 
domain, as well as various technologies being taken into account by the companies
operating in this industry.
We have proposed a novel mixed-criticality multi-OS architecture based on open hardware and open-source software. We have then described the optimizations done both at the software and hardware levels
to move performance closer to the commercial competitors.
The experimental results have shown performance comparable to the state-of-the-art and have also allowed identifying further room for future hardware optimizations of 
the CVA6 RISC-V processor.
In the future, we plan further to evolve the proposed architecture by (i) designing advanced hardware features such as banked stack pointers and optimized context switch to improve the competitiveness of the CVA6 architecture further, (ii) leveraging the recent standardization of DDS in Classic AUTOSAR~\cite{rti-dds} by using plain DDS instead of DDS-XRCE for inter-domain communications, and (iii) adopting the \texttt{SCHED\_DEADLINE} scheduling policy~\cite{lelli2016} to have a more predictable timing behavior of the communication on the Linux OS.

\bibliographystyle{IEEEtran}
\bibliography{bibliography}

\end{document}